# Recursive Threshold Median Filter and Autoencoder for Salt-and-Pepper Denoising: SSIM analysis of Images and Entropy Maps

Petr Boriskov[a, *], Kirill Rudkovskii[a], and Andrei Velichko[a]

[a]*Petrozavodsk State University, Petrozavodsk, Russia*
*e-mail: boriskov@petrsu.ru

**Abstract**
This paper studies the removal of salt-and-pepper noise from images using median filter (MF) and simple three-layer autoencoder (AE) within recursive threshold algorithm. The performance of denoising is assessed with two metrics: the standard Structural Similarity Index $SSIM_{Img}$ of restored and clean images and a newly applied metric $SSIM_{Map}$ — the SSIM of entropy maps of these images computed via 2D Sample Entropy in sliding windows. We shown that $SSIM_{Map}$ is more sensitive to blur and local intensity transitions and complements $SSIM_{Img}$. Experiments on low- and high-resolution grayscales images demonstrate that recursive threshold MF robustly restores images even under strong noise (50-60 %), whereas simple AE is only capable of restoring images with low levels of noise (<30 %). We propose two scalable schemes: (i) 2MF, which uses two MFs with different window sizes and a final thresholding step, effective for highlighting sharp local details at low resolution; and (ii) MFs-AE, which aggregates features from multiple MFs via an AE and is beneficial for restoring the overall scene structure at higher resolution. Owing to its simplicity and computational efficiency, MF remains preferable for deployment on resource-constrained platforms (edge/IoT), whereas AE underperforms without prior denoising. The results also validate the practical value of $SSIM_{Map}$ for objective blur assessment and denoising parameter tuning.

## 1 Introduction

Over the past decade, a wide range of image denoising algorithms has been proposed, offering different quality levels and computational costs. Transform-domain methods—wavelet-based [1,2] and spectral/frequency approaches [3], singular-value–based techniques [1,4], and three-dimensional filtering [5]—are effective at suppressing noise while preserving useful content. However, these approaches are often complex and require substantial domain expertise due to numerous variants and parameters. For example, the choice of the mother wavelet and the decomposition level strongly affects performance. In addition, deploying computationally demanding algorithms on low-power or embedded platforms can be challenging.

Local filters such as the moving-average and the MF [6] are long-established, offering very high throughput at low computational cost. In their simplest forms, though, they provide limited tunability—essentially only the sliding-window (kernel) size. As the kernel grows, stronger denoising is accompanied by increased blurring and reduced contrast.

Impulse noise—often called salt-and-pepper (SP) noise in images—remains a major challenge in digital image processing [6,7]. Among common filters, the MF is one of the most effective for SP noise removal. The MF is a nonlinear operator: pixels within each sliding window are ranked, and the central order statistic is returned. Unlike the mean filter, it suppresses extreme outliers without greatly affecting neighboring values. Because edges are largely preserved, the median filter can be applied recursively. The advent of recursive MFs [8,9] further improved the denoising efficacy of median-based schemes. To mitigate blurring, the MF is often combined with thresholding [10]: the filtered image is compared with the original, and original pixel values are retained where a specified threshold condition is satisfied.

Over the past 30–40 years, advances in artificial neural networks have driven the development of denoising algorithms built on a variety of architectures—from feedforward networks (multilayer perceptrons) [11] and recurrent schemes [12] to convolutional networks [13,14] and deep-learning models [15–18]. Neural AE are a staple of modern deep learning [16–18]. Their primary role in data processing is to learn compact, salient representations (features), which is particularly important for recognition/classification, clustering, and related tasks. Low-salience components are typically attenuated or removed and thus do not propagate further through the pipeline. Because representation compression inherently suppresses background noise, autoencoders naturally act as denoisers. Denoising AEs can serve as lightweight preprocessing modules that improve performance both in cloud-hosted deep models [16,17] and in on-device/edge computing for mobile IoT scenarios (microcontrollers, sensors) [17,18].

In this work, we perform a detailed analysis of recursive threshold removal of SP corruption in grayscale images using denoising filters (hereafter, DnF): MF and a simple three-layer AE. We insert a threshold rule into the output-to-input feedback loop (recursing) of both DnF: the pixels of output image are either retained or replaced with the corresponding pixels from the original noisy image. Building on this recursing, we propose a scalable, multi-level recursive threshold algorithm for restoring heavily corrupted SP images. The algorithm can employ different sizes of sliding windows and threshold levels in parallel (2MF) and parallel-serial (MFs-AE) noise reduction schemes. In all schemes, the input is a noisy image in one copy (pattern); the clean image (reference) is used solely for evaluating denoising quality.

As the quality metric of image restoring, we use the standard SSIM metric [19]. A key novelty of our approach is the systematic use of SSIM not only on the images themselves but also on their entropy maps, which we obtain by computing two-dimensional Sample Entropy (2D SampEn) [20,21] over sliding windows.

The rest of the paper is organized as follows. Section 2 describes the recursive denoising scheme with a threshold rule, details of the AE as an machine learning algorithm, and the SSIM computation for images and their entropy maps. Section 3 (Results) reports restorations for grayscale images at low and high resolutions: a white mushroom (*porcini*, 100×150 px) and *Lena* (512×512 px). A comparative analysis of MF vs. AE is given in Section 3.1, while the scalable algorithm for the 2MF and MFs-AE schemes appears in Sections 3.2 and 3.3. Sections 4 (Discussion) and 5 (Conclusion) summarize the proposed algorithms. All computations were performed in Python.

## 2 Methods

### 2.1 SP denoising by MF and AE with recursive thresholding

The general DnF scheme with recursive thresholding is shown in Fig. 1a, where the recursing means that the once-filtered image is fed back to the input. The initial noisy image, hereinafter referred to as the source image, is fed to the DnF input with SP corruption and normalized grayscale intensity in the range [0, 1] with 0 (white) and 1 (black). SP noise is modeled as uncorrelated pixelwise perturbations randomly scattered across the reference grid and is commonly specified in two forms [22]: interval and fixed SP noise. In this study, the primary model is interval SP noise with amplitudes uniformly distributed in two narrow bands: [0, 0.1] ("pepper") and [0.9, 1] ("salt"). The fixed model, where "salt" and "pepper" are exactly 1 and 0, is used only for benchmarking against established denoising methods (see Table 4). The SP noise level $\mu_{sp}$ is defined, for both models, as the fraction of corrupted pixels:

$$\mu_{sp}(\%) = 100 \cdot \frac{n_c}{n}, \tag{1}$$

where $n_c$ is the combined number of corrupted "salt" and "pepper" pixels (assumed equal density), and $n$ is the total number of pixels in the image.

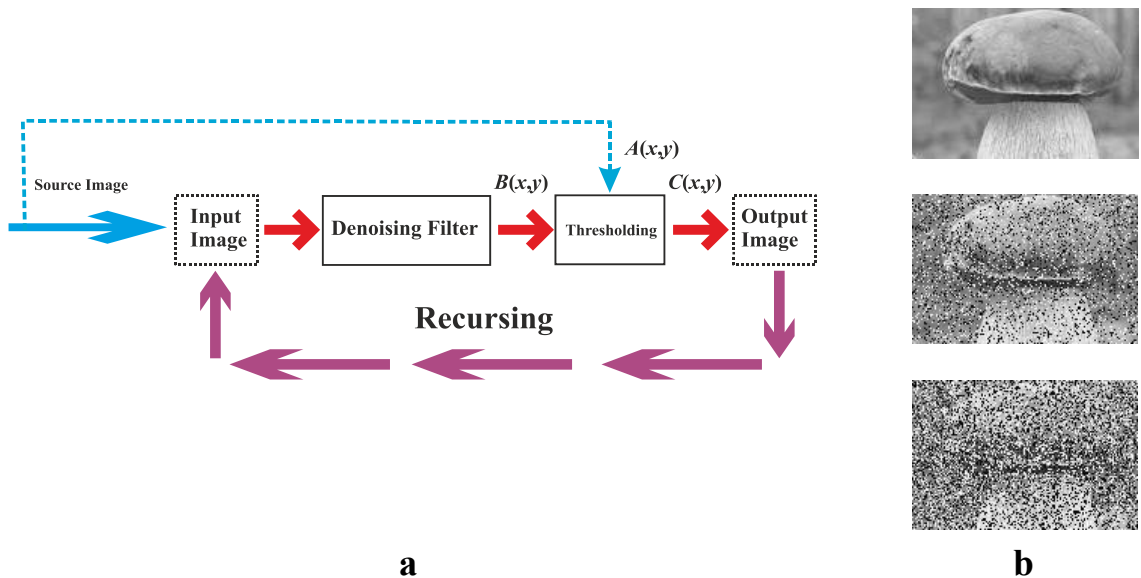

**Fig. 1.** (a) General scheme of denoising filter (MF or AE) with recursive thresholding. The source image is used only at the first pass (blue solid arrow) and in thresholding (2) (blue dashed arrow). $A(x, y)$, $B(x, y)$, and $C(x, y)$ denote pixels of the source, DnF output, and post-threshold images, respectively. (b) The example of porcini image (100×150 px): clean (top) and noisy versions with $\mu_{sp}$ = 23.3 % (middle) and $\mu_{sp}$ = 46.6 % (bottom).

**Threshold rule.** In both DnF scenarios, a threshold rule is inserted into the output-to-input feedback loop: depending on this rule, output pixels are either retained or replaced by the corresponding pixels from the source image. For every pixel the thresholding is defined as

$$C(x,y) = \begin{cases} A(x,y), & \text{if } |A(x,y) - B(x,y)| < threshold \\ B(x,y), & \text{if } |A(x,y) - B(x,y)| > threshold \end{cases}, \quad (2)$$

where $A(x,y)$, $B(x,y)$ and $C(x,y)$ are grayscale intensities [0,1] at coordinates $(x,y)$ for the source, the filter output, and the post-threshold images, respectively (see Fig. 1a). Thus, at each DnF pass, output pixels are replaced by the corresponding pixels of initial noisy (source) image when the absolute difference $A(x,y)$ and $B(x,y)$ does not exceed the threshold. Evidently, with a zero threshold the rule degenerates to no thresholding, i.e., $C(x,y) = B(x,y)$ when $threshold = 0$. Equation (2) will be modified for scaled denoising (Sections 3.2–3.3); see Section 4 (Discussion) for analysis.

**MF.** Within each primary block, denoising is performed by a sliding window; the median in the window $W$ over pixels $A(x,y)$ is computed as [7]

$$median = \underset{A(x_o, y_o) \in W(x,y)}{\arg\min} \sum |A(x,y) - A(x_o, y_o)|. \quad (3)$$

Rectangular windows symmetric about the center may be (3×3), (5×5), (7×7), (9×9) etc.

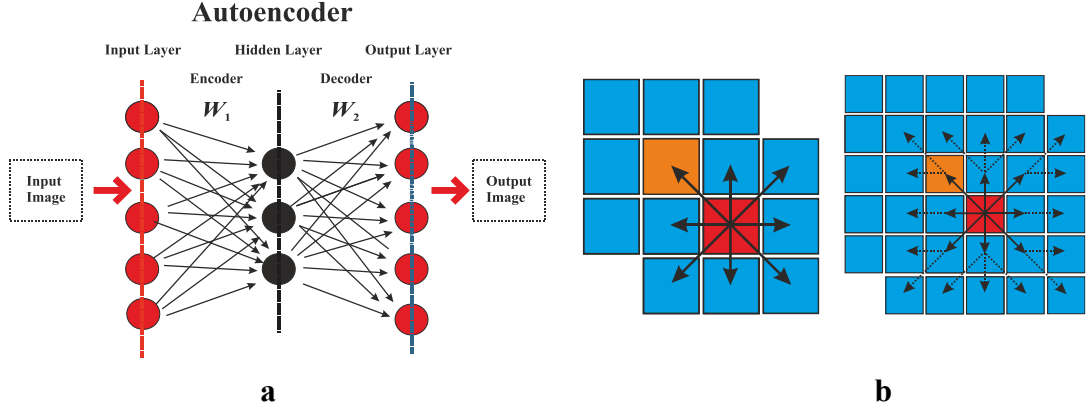

**Fig.2.** (a) Architecture of three-layer AE. $W_1$ and $W_2$ denote the encoder and decoder weight matrices; sigmoid activation is used in the hidden and output layers neurons. (b) Training-set generation scheme for left (3×3) and right (5×5) windows: solid arrows indicate a one-step shift of the central pixel; dashed arrows (for the (5×5) window) indicate an additional step. The window with the central pixel (red square) is shifted toward the upper-left corner; the new position is shown in orange.

**AE.** We adopt block-wise processing to speed up computation and reduce memory in the AE algorithm (Fig. 2a): the image is partitioned into equal, non-overlapping blocks, and each block is denoised independently. Before training each block, the encoder and decoder weight matrices are re-initialized at random. We refer to these partitioning units as primary blocks.

The same symmetric sliding windows are used to construct the training set of the AE. The primary block size $N_{bl}$ determines the number of input/output layers neurons (Fig. 2a); e.g., (50×50) block yields 2500 neurons. We form a training set consisting of the primary block and its shifts in all directions by a specified number of pixel steps. Fig. 2b illustrates directions and set size using the same window family as for MF. With (3×3) window, there are 9 patterns (the block plus all one-step shifts along axes and diagonals). With (5×5) window, shifts by one and two steps are included (25 patterns). For (7×7) window, there are 49 patterns (one-, two-, and three-step shifts). Thus, larger windows incorporate pixels progressively farther from the primary block. In general, for ($m \times m$) window: $N_{learn} = m^2$.

## 2.2 SSIM of images and their entropy maps

A standard measure of image similarity is the Structural Similarity Index (SSIM), which compares two images via local-window statistics: means, variances, and covariance of pixel intensities. The combined local metric $SSIM_{local}$ is [19]:

$$SSIM_{\text{local}}(x, y) = \frac{(2\mu_x\mu_y + C_1)(2\sigma_{xy} + C_2)}{(\mu_x^2 + \mu_y^2 + C_1)(\sigma_x^2 + \sigma_y^2 + C_2)}, \qquad (4)$$

where $\mu_x$ ($\mu_y$) and $\sigma_x^2$ ($\sigma_y^2$) are the means and variances in the corresponding local windows of the two images, $\sigma_{xy}$ is their covariance (the window location is $(x, y)$, $C_1 = (0.01\cdot 255)^2$ and $C_2 = (0.03\cdot 255)^2$ are empirical constants tied to the intensity range that ensure numerical stability. The global SSIM for an image pair is the arithmetic mean of $SSIM_{\text{local}}$ over all window positions.

For any raster image, one can define families of overlapping local windows that slide with stride 1; the families differ only by window size, and the number of window positions equals the number of image pixels. Using any such family, we construct an entropy map of the same size by computing for each pixel of the image the entropy of its sliding window (e.g., 2D Sample Entropy) and assigning it to that pixel (normalized to [0,1]). A single image admits multiple entropy maps depending on window size and entropy parameters; each map captures a feature set specific to only one particular image.

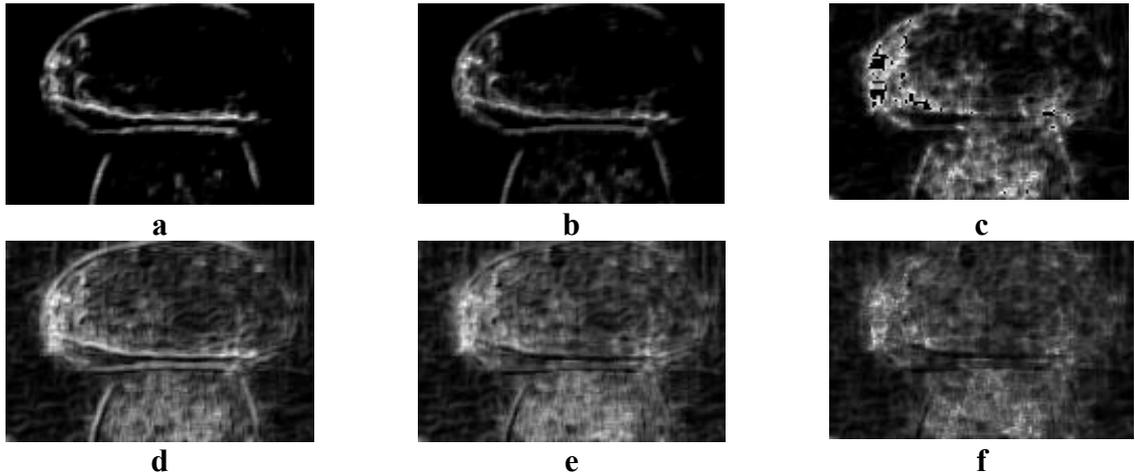

**Fig. 3.** Entropy maps of the clean mushroom image (Fig. 1b, top) computed with (5×5) sliding windows: SampEn [21] with tolerance $r = 0.2$ (a), $r = 0.15$ (b) and $r = 0.05$ (c); FuzzyEn [23] with $n = 2$ and $r = 0.0038$ (d), $r = 0.001$ (e) and $r = 0.0002$ (f). Embedding dimension $m=1$ and sampling time $\tau = 1$ are in SampEn and FuzzyEn.

Figure 3 illustrates multiple entropy maps of the same clean image (Fig. 1b, top) computed using two well-known 2D entropy measures—Sample Entropy (SampEn) [21] and Fuzzy Entropy (FuzzyEn) [23]. The maps differ markedly in segmentation and texture. Since entropy will be used as an auxiliary blur indicator for restored images, we adopt parameters that yield the sharpest boundary delineation; in what follows we use SampEn with the parameters of Fig. 3a: embedding dimension $m=1$, sampling time $\tau = 1$, tolerance interval $r = 0.2$.

When two entropy maps share the same size, their similarity can be assessed exactly as for real images: via local $SSIM_{\text{local}}$ (4) and the global SSIM (the mean of $SSIM_{\text{local}}$ over all window positions). In the denoising task, we compute entropy maps of the clean and restored images and take the global SSIM as their average over sliding windows. This new parameter is an additional characteristic of the similarity between clean and restored images, which can be used to evaluate the effectiveness of noise reduction. To distinguish metrics, let $SSIM_{\text{Img}}$ denote SSIM for images and $SSIM_{\text{Map}}$ for entropy maps. We also use for SSIM metrics the relative difference in percent:

$$\Delta SSIM\,(\%) = 100 \cdot \frac{|SSIM_1 - SSIM_2|}{(SSIM_1 + SSIM_2)/2}, \qquad (5)$$

where $SSIM_1$ and $SSIM_2$ are either $SSIM_{\text{Img}}$ and $SSIM_{\text{Map}}$.

## 3 Results

**3.1 Comparative analysis of SP denoising by MF and AE with recursive thresholding**

Table 1 lists the DnF (MF and AE) parameters. Some values vary by experiment: the noise level in Figs. 4–5 and the threshold in Fig. 6a. Figure 4 illustrates the effect of the threshold rule (2) on SP denoising in terms of SSIM metrics ($SSIM_{\text{Img}}$ and $SSIM_{\text{Map}}$). For both filters (MF and AE), these metrics are higher with thresholding (Fig. 4b) than without (Fig. 4a). Moreover, as seen from the blue curve with circles in Fig. 4a (MF, $SSIM_{\text{Img}}$), even at very low noise $\mu_{\text{sp}} \leq 10\,\%$ the values do not exceed ~ 0.81, whereas the SSIM metrics with thresholding are close to 1 (Fig. 4b). Note that images with $SSIM_{\text{Img}} > 0.9$ are visually almost indistinguishable from the reference (Fig. 1b, top), while for $SSIM_{\text{Img}} < 0.2$ effective restoration is essentially absent.

Figure 4 illustrates the effect of the threshold rule (2) on SP denoising in terms of SSIM metrics ($SSIM_{Img}$ and $SSIM_{Map}$). For both filters (MF and AE), these metrics are higher with thresholding (Fig. 4b) than without (Fig. 4a). Moreover, as seen from the blue curve with circles in Fig. 4a (MF, $SSIM_{Img}$), even at very low noise $\mu_{sp} \leq 10\,\%$ the values do not exceed ~ 0.81, whereas the SSIM metrics with thresholding are close to 1 (Fig. 4b). Note that images with $SSIM_{Img} > 0.9$ are visually almost indistinguishable from the reference (Fig. 1b, top), while for $SSIM_{Img} < 0.2$ effective restoration is essentially absent.

**Table 1** DnF parameters in Section 3.1

| DnF | block shape, $N_{bl}$ | window shape, $N_w$ | noise level, $\mu_{sp}$ (%) | | | threshold | epochs | learning rate | compression ratio |
|---|---|---|---|---|---|---|---|---|---|
| | | | Fig. 6a | Fig. 6b | Fig. 7 | | | | |
| MF | – | (5×5) | 23.3 | 66.6 | 46.6 | 0.2 | – | – | – |
| AE | (50×50) | | | | | | 20 | 0.001 | 0.5 |

\* The block shape in the MF is equal full size image (100×150, $n$=15000 px). The compression ratio of AE is the ratio of the number of neurons in the input / output layer to the number of neurons in the hidden layer.

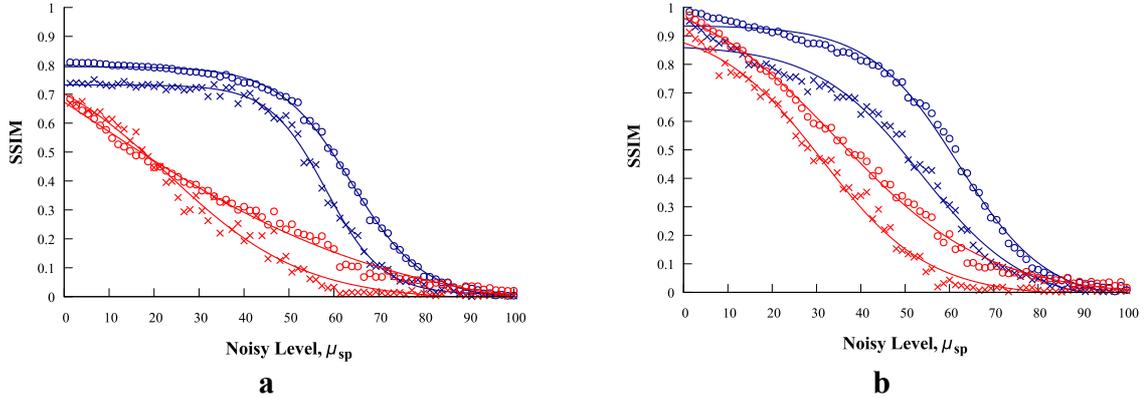

**Fig. 4.** Dependence of SSIM metrics ($SSIM_{Img}$ – circles and $SSIM_{Map}$ – crosses) on the SP noise level $\mu_{sp}(\%)$ for non-recursive (single-pass) denoising by MF (blue) and AE (red): without threshold (a) and with threshold (b). Calculation parameters are in Table 1.

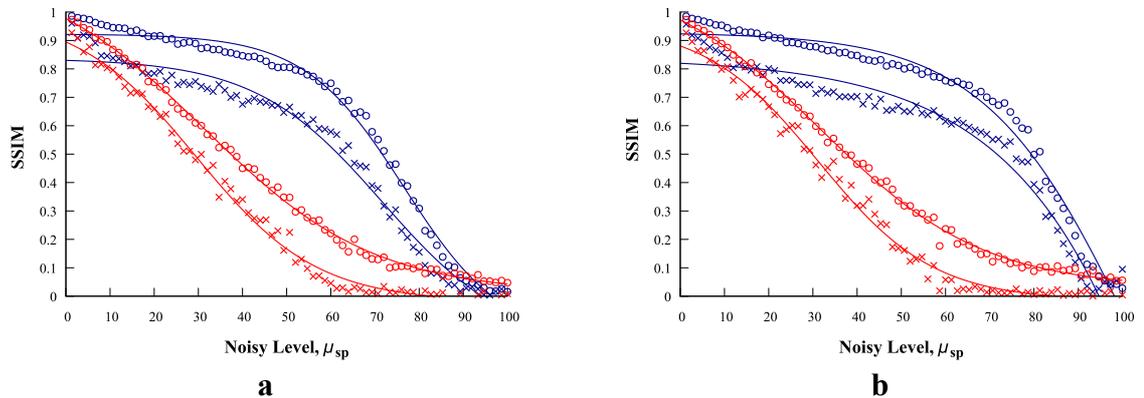

**Fig. 5.** Dependence of SSIM metrics ($SSIM_{Img}$ – circles and $SSIM_{Map}$ – crosses) on the SP noise level $\mu_{sp}(\%)$ for denoising by MF (blue) and AE (red) after one (a) and five (b) recursions. Calculation parameters are in Table 1.

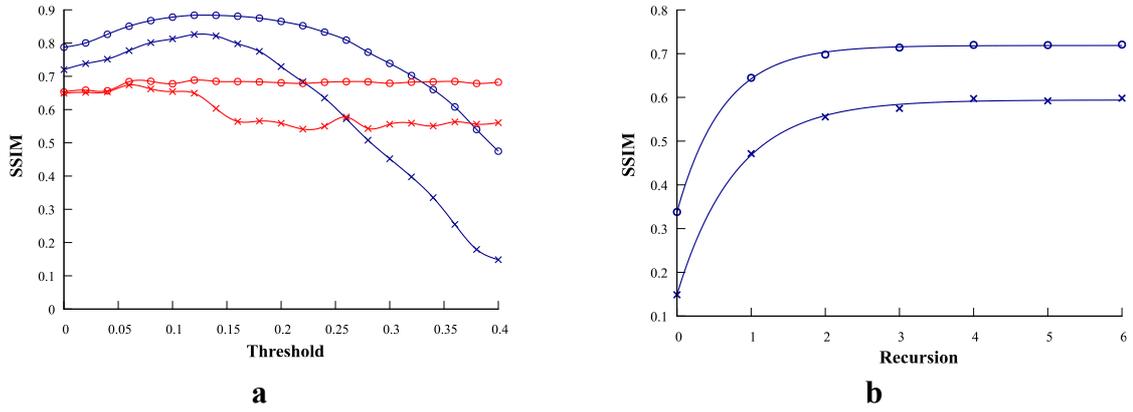

**Fig. 6.** Dependence of SSIM metrics ($SSIM_{Img}$ – circles and $SSIM_{Map}$ – crosses) on the threshold (a) and on the number of recursions (b) for denoising by MF (blue) and AE (red). Calculation parameters are in Table 1.

The introducing of recursive processing (Fig. 5) has little effect on denoising efficiency at low noise levels for both MF and AE. As $\mu_{sp}$ increases, however, MF improves markedly, whereas AE remains almost unchanged relative to the non-recursive case (cf. Fig. 4b). Up to very high noise ($\mu_{sp} \geq 70$ %) the median filter yields substantially higher $SSIM_{Img}$ and $SSIM_{Map}$ than the autoencoder, especially with recursive thresholding (Fig. 5). The curves $SSIM_{Img}(\mu_{sp})$ and $SSIM_{Map}(\mu_{sp})$ (Figs. 4–5) are nonlinear and well fitted by exponential sigmoids. As the number of recursions increases, the MF sigmoids (Fig. 5) become slightly broader and the noise level at which the curve crosses 0.5 (midpoint/inflection) shifts to the right relative to the blue curves in Fig. 4b. Note that $SSIM_{Map}$ is consistently below $SSIM_{Img}$ at the same noise level. The dependencies on the threshold (Fig. 6a) exhibit a clear maximum for MF (blue curves). For AE, the threshold has little effect at low noise (red curves in Fig. 6a) and degrades performance at higher $\mu_{sp}$.

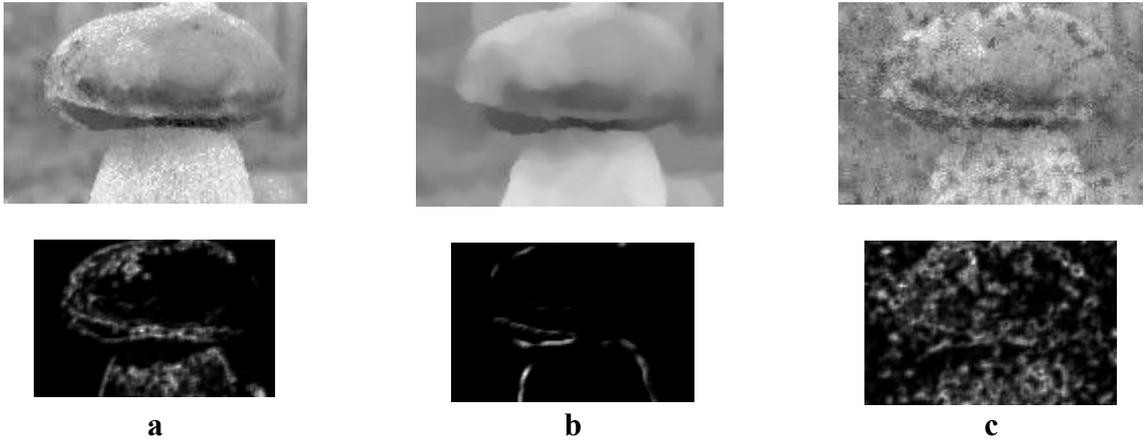

**Fig. 7.** Restored images (top) of the heavily SP corrupted image (Fig. 1b, bottom, $\mu_{sp} = 46.6$ %) after 10 recursions and their entropy maps (bottom) obtained using: (a) MF with thresholding; (b) MF without thresholding; (c) AE with thresholding. SSIM metrics: (a) $SSIM_{Img} = 0.82$, $SSIM_{Map} = 0.64$; (b) $SSIM_{Img} = 0.72$, $SSIM_{Map} = 0.58$; (c) $SSIM_{Img} = 0.39$, $SSIM_{Map} = 0.12$. Calculation parameters are in Table 1.

Numerically, MF shows weak sensitivity to recursion at low–moderate $\mu_{sp}$. However, as Figure 6b shows, under very strong corruption ($\mu_{sp} > 60$ %), $SSIM_{Img}$ and $SSIM_{Map}$ jump sharply after the first recursion and then quickly (within a few passes) approach slowly increasing quasi-stationary values; full stabilization typically occurs by ~10–15 passes. In contrast, AE quality is almost insensitive to the number of recursions at any noise level.

As seen in Fig. 7a (top), increasing the number of recursions enables effective suppression of strong SP noise (Fig. 1b, bottom) by MF. In contrast, AE performs poorly (Fig. 7c). The introducing of threshold rule (2) into the MF recursion markedly improves denoising, as evident from comparing Figs. 7a and 7b: without thresholding, the restored image is strongly blurred (Fig. 7b, top) and the mushroom boundary is weak on the entropy map (Fig. 7b, bottom). The computed $SSIM_{Img}$ and $SSIM_{Map}$ (see the caption of Fig. 7) corroborate these observations: the best similarity to the clean image is achieved by MF with thresholding, and the worst by AE. Note also that for Fig. 7 the relative difference (5) between $SSIM_{Map}$ values is larger than that between $SSIM_{Img}$ values.

## 3.2 Two-level Scaled SP Denoising Based on the 2MF Scheme

**Table 2** MF parameters in Section 3.2

| MF | Threshold | | | | | | |
|---|---|---|---|---|---|---|---|
| | Fig. 8 (100×150) | | Fig. 9 (2MF) (100×150) | | Fig. 10 (40×50) | | |
| | Fig.8b | Fig. 8c | Step 1 | Step 2 | Fig.10b | Fig. 10c (2MF) | |
| | | | | | | Step 1 | Step 2 |
| MF $W_1$ (3×3) | 0.1 | – | 0.1 | 0.2 | – | 0.1 | 0.2 |
| MF $W_2$ (5×5) | – | 0.1 | | | 0.1 | | |

\* The size image is (100×150) px in Fig. 8 and 9, the size image is (40×50) px in Fig.10, the number of recursion in Fig.8-10 is 25.

Figure 8 shows how strongly the sliding-window size affects MF denoising at very high corruption levels ($\mu_{sp}$ > 60 %, $n_c$ > 9000). As can be seen, the pixels in the restored image, which were initially scattered uniformly (Fig. 8a, top), are combined into "islands" (Fig. 8a, bottom). These clusters of impulse artifacts cannot be removed merely by increasing the number of recursions while keeping the same window size (3×3). Enlarging the window to (5×5) eliminates the islands—the output image contains virtually no impulse corruption (Fig. 8b, top). The trade-off is a moderate loss of sharpness in the restored image. Further increases in window size provide stronger noise suppression but at the cost of additional blurring.

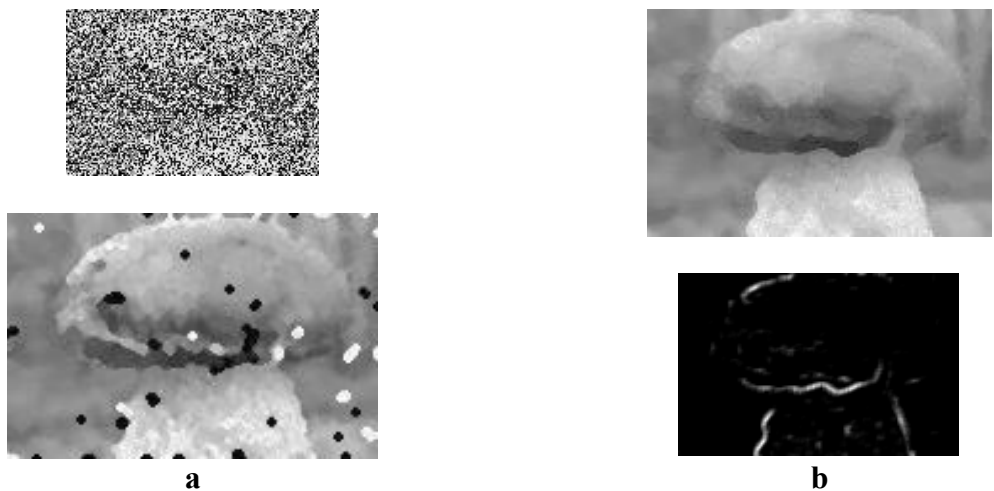

          **a**                                                  **b**

**Fig. 8.** (a) SP heavily corrupted image at $\mu_{sp}$ = 66.6 % (top) and its restoration by MF with (3×3) window (bottom): $SSIM_{Img}$ = 0.63. (b) Restored image (top) and its entropy map (bottom) obtained by the MF with (3×3) window: $SSIM_{Img}$ = 0.724, $SSIM_{Map}$ = 0.61. Calculation parameters are in Table 2.

We propose a promising two-scale scheme that combines two median filters with window sizes $W_1 < W_2$ (Fig. 9a), the algorithm of which is in two steps:
*Step 1.* Apply recursive threshold denoising (multiple passes) to the noisy image with two independent MFs using windows $W_1$ and $W_2$ as in Fig. 1a, yielding Out Img $W_1$ and Out Img $W_2$.
*Step 2 –Thresholding.* Use rule (2) with $A(x, y)$ is Out Img $W_1$, $B(x, y)$ is Out Img $W_2$, and $A(x, y)$ is the final Out Img. We refer to this 2MF scheme as a two-level scaled recursive median algorithm with thresholding, since two window scales are combined.

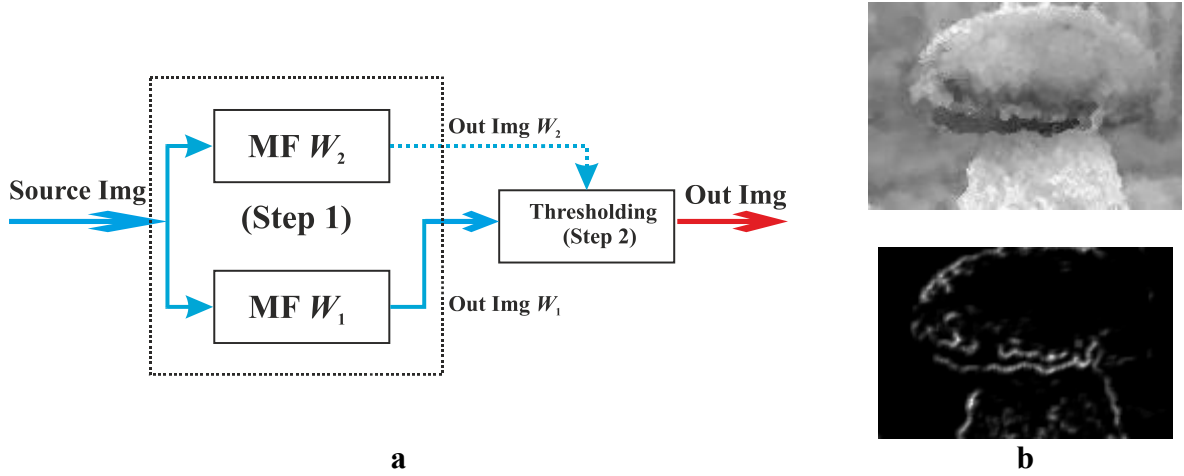

**a** **b**

**Fig. 9.** (a) 2MF denoising scheme for heavily SP corrupted images. (b) Restored image (top) and its entropy map (bottom) obtained by the 2MF scheme from SP heavily corrupted image (Fig. 8a, $\mu_{sp}$ = 66.6 %): $SSIM_{Img}$ = 0.736, $SSIM_{Map}$ = 0.66. Calculation parameters are in see Table 2.

Fig. 9b shows the restoration of a heavily damaged SP image by the 2MF scheme with $W_1$ = (3×3) and $W_2$ = (5×5) and two thresholds at Steps 1 and 2 (Table 2). The relative difference $\Delta SSIM_{Img}$ (5) for 2MF scheme exceeds the single-window result (Fig. 8b, top) by only ≈ 1.5%, whereas $\Delta SSIM_{Map}$ increases by nearly 8%. Thus, SSIM-based assessments, especially $SSIM_{Map}$, confirms that 2MF output image is slightly less blurred than the restored image of single-window MF.

Figure 10 further shows denoising of a region with size (40×50) px containing a sharp edge (the mushroom cap). The 2MF restoration in Fig. 10c yields $\Delta SSIM_{Img}$ and $\Delta SSIM_{Map}$ larger by about 8% and 27%, respectively, compared with the single-window MF (Fig. 10b).

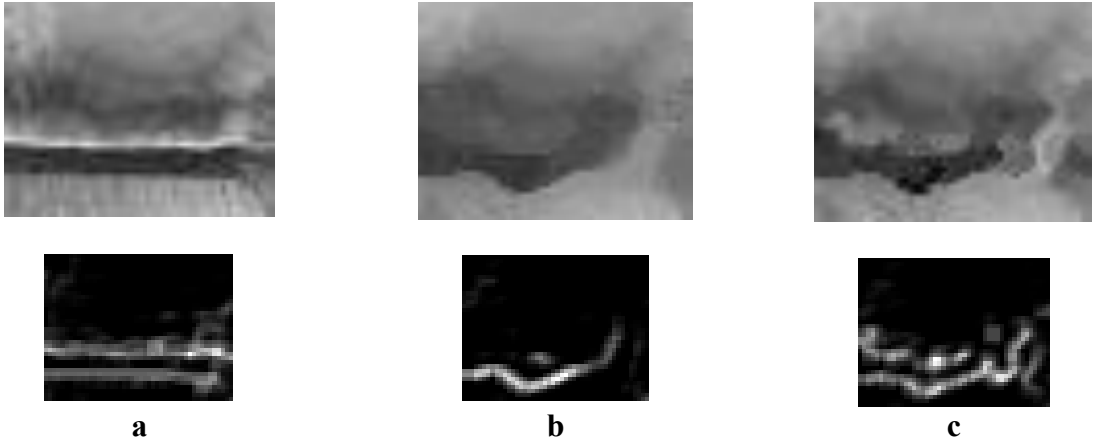

**a** **b** **c**

**Fig. 10.** (a) A patch of the clean image (Fig. 1b, top) containing the mushroom-cap edge (top) and its entropy map (bottom). (b) and (c) Restored patches (top) of the heavily corrupted image (Fig. 8a, $\mu_{sp}$ = 66.6 %) and their entropy maps (bottom) obtained with single-window MF (5×5) and the 2MF scheme, respectively. SSIM metrics: (b) $SSIM_{Img}$ = 0.61, $SSIM_{Map}$ = 0.41; (c) $SSIM_{Img}$ = 0.66, $SSIM_{Map}$ = 0.54. Calculation parameters are in see Table 2.

These results (especially $SSIM_{Map}$) indicate that fine details are more clearly visible by the scalable 2MF scheme (Fig. 9a). Indeed, the cap edge on the 2MF entropy map (Fig. 10c, bottom) is sharply delineated as a region with a double intensity transition (see Fig. 10a, bottom), whereas it is largely absent in Fig. 10b (bottom). The relative difference $\Delta SSIM_{Map}$ (5) between the maps of Fig. 10b and 10c (bottom), which is about 27 %, confirms the high sensitivity of $SSIM_{Map}$ to the presence/absence of sharp details in zoomed, low-resolution patches.

### 3.3 Multi-level Scaled SP Denoising Based on the MFs-AE Scheme

Figure 11a presents an extended denoising scheme that augments 2MF scheme with AE. The algorithm proceeds as follows.
*Step 1.* The source image (Source Img) is filtered independently by a single MF with window $W_2$ and the set of MFs with window $W_1$ ($W_1 < W_2$):

$$\{nMF\ W_1\} = \{MF_1\ W_1, MF_2\ W_1, \ldots, MF_n\ W_1\}. \tag{6}$$

MFs follow the general scheme of Fig. 1a, that is, they denoise the source image with several threshold-based recursions. The thresholds in set $\{nMF\ W_1\}$ (6) are in the range $\{threshold_{min}, threshold_{max}\}$ with step $\Delta thr$; MF $W_1$ uses one fixed threshold from the same range. The set of filters $\{nMF\ W_1\}$ and MF $W_2$ yields output images (Fig 11a): $\{Out_1, Out_2, \ldots Out_n\}$ and Out Img $W_2$, respectively.

*Step 2.* The set of output images $\{Out_1, Out_2, \ldots Out_n\}$ are fused into a single aggregate image Out AE by the autoencoder. The training set of AE $\{Input\ AE\}$ (Fig. 11a) is constructed from $\{Out_1, Out_2, \ldots Out_n\}$ using the same sliding-window procedure as for a single image (see Fig. 2b). For an ($m \times m$) window and $n$ images, the number of training patterns is $N_{learn} = n \cdot m^2$. Note that with the minimal (1×1) window (single pixel), $N_{learn} = n$, that is, simply equal to the number of output images of the set of filters $\{nMF\ W_1\}$. We further use the minimum window in our calculations (see the heading of Table 3).

**Table 3** The parameters of MFs-AE scheme in Section 3.3

| DnF | Fig.11, (b) – the size image is (100×150) px, (c) – the size image is (40×50) px, Fig.12 and Table 4 – the size image is (512×512) px | | | | | | |
|---|---|---|---|---|---|---|---|
| | $threshold_{min}$ | $threshold_{max}$ | $\Delta thr$ | threshold | epochs | learning rate | compression ratio |
| MF $W_1$ (3×3), $n$ = 8 | 0.08 | 0.15 | 0.01 | – | – | – | – |
| MF $W_2$ (5×5) | – | – | – | 0.1 | – | – | – |
| AE (1×1) | – | – | – | 0.25 | 100 | 0.001 | 0.4 |

\* The number of recursion in MF $W_1$ and MF $W_2$ is 25, the block shape $N_{bl}$ in AE is (50×50) and the training set window is (1×1) = 1 px, that is $N_{learn} = n = 8$.

Figure 11a presents an extended denoising scheme that augments 2MF scheme with AE. The algorithm proceeds as follows.

*Step 1.* The source image (Source Img) is filtered independently by a single MF with window $W_2$ and the set of MFs with window $W_1$ ($W_1 < W_2$):

$$\{nMF\ W_1\} = \{MF_1\ W_1, MF_2\ W_1, \ldots, MF_n\ W_1\}. \tag{6}$$

MFs follow the general scheme of Fig. 1a, that is, they denoise the source image with several threshold-based recursions. The thresholds in set $\{nMF\ W_1\}$ (6) are in the range $\{threshold_{min}, threshold_{max}\}$ with step $\Delta thr$; MF $W_1$ uses one fixed threshold from the same range. The set of filters $\{nMF\ W_1\}$ and MF $W_2$ yields output images (Fig 11a): $\{Out_1, Out_2, \ldots Out_n\}$ and Out Img $W_2$, respectively.

*Step 2.* The set of output images $\{Out_1, Out_2, \ldots Out_n\}$ are fused into a single aggregate image Out AE by the autoencoder. The training set of AE $\{Input\ AE\}$ (Fig. 11a) is constructed from $\{Out_1, Out_2, \ldots Out_n\}$ using the same sliding-window procedure as for a single image (see Fig. 2b). For an ($m \times m$) window and $n$ images, the number of training patterns is $N_{learn} = n \cdot m^2$. Note that with the minimal (1×1) window (single pixel), $N_{learn} = n$, that is, simply equal to the number of output images of the set of filters $\{nMF\ W_1\}$. We further use the minimum window in our calculations (see the heading of Table 3).

*Step 3.* Apply the thresholding rule (2) to compare Out AE with Out Img $W_2$, similar to the 2MF scheme (Fig.9a): $A(x, y)$ = Out AE, $B(x, y)$ = Out Img $W_2$, $C(x, y)$ = Out Img is the final output. We refer to this scheme MFs-AE as a multi-level scaled recursive median algorithm with thresholding, since it uses multiple MFs.

Figures 11b and 11c show restored images and their entropy map using MFs-AE (Fig. 11a) with $W_1$ = (3×3) and $W_2$ = (5×5). For the full image (Fig 11b), the relative differences (5) yield ~3% for $SSIM_{Img}$ and ~12% for $SSIM_{Map}$ higher than the single-window MF (Fig. 8b, top). This is a little better than for the 2MF scheme in Fig. 9b: ~1.5% and ~8%, respectively. However, for the patch 40 × 50px (mushroom-cap edge), MFs-AE shows no advantage over the single-window MF: $SSIM_{Img}$ in Fig. 10b (top) and Fig. 11c (top) are nearly identical (~0.61 vs 0.614), while $SSIM_{Map}$ = 0.39 in Fig. 11c (bottom) is even lower than 0.41 in Fig. 10b (bottom).

We also consider the well-known *Lena* image (512×512 px) at higher resolution. Figure 12 shows the clean image, the heavily corrupted version, and the restorations using single-window MFs with (3×3) and (5×5), as well as with the MFs-AE scheme (Fig. 11a). Figure 13 presents the corresponding entropy maps in two variants: standard 2D Sample Entropy maps (left) and their contrast-enhanced copies (right) produced via morphological dilation [7]. Note that, for this case, the 2MF scheme (Fig. 9a) yields a result practically indistinguishable from MF with (5×5) window (Fig. 12c) and is therefore omitted.

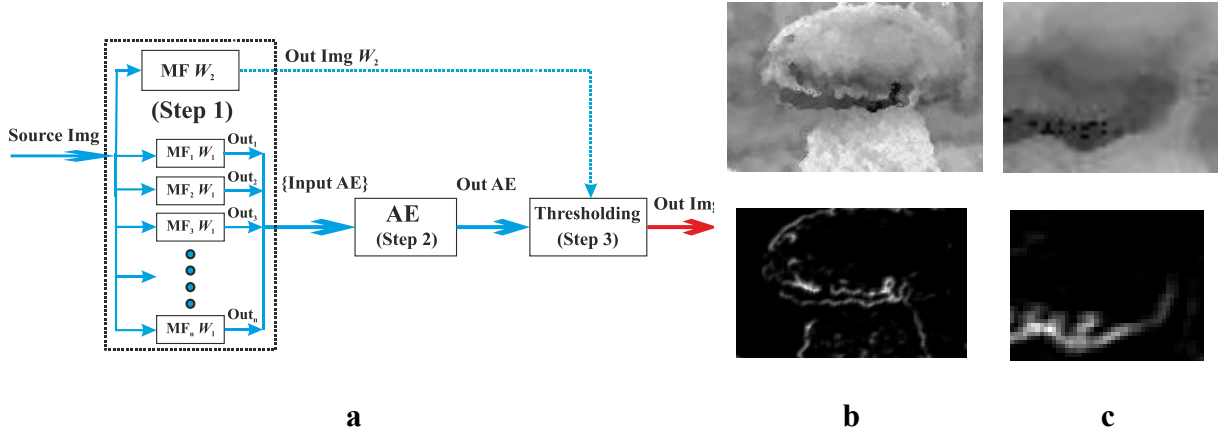

**Fig. 11.** (a) MFs-AE denoising scheme for heavily corrupted images. (b) and (c) Restored images (top) and their entropy maps (bottom) obtained by the MFs-AE scheme for the source image with $\mu_{sp}$ = 66.6 % (Fig. 8a): the full frame (100×150) and (40×50) px patch (mushroom-cap edge), respectively. SSIM metrics: (b) $SSIM_{Img}$ = 0.745, $SSIM_{Map}$ = 0.69; (c) $SSIM_{Img}$ = 0.614, $SSIM_{Map}$ = 0.39. Calculation parameters are in Table 3.

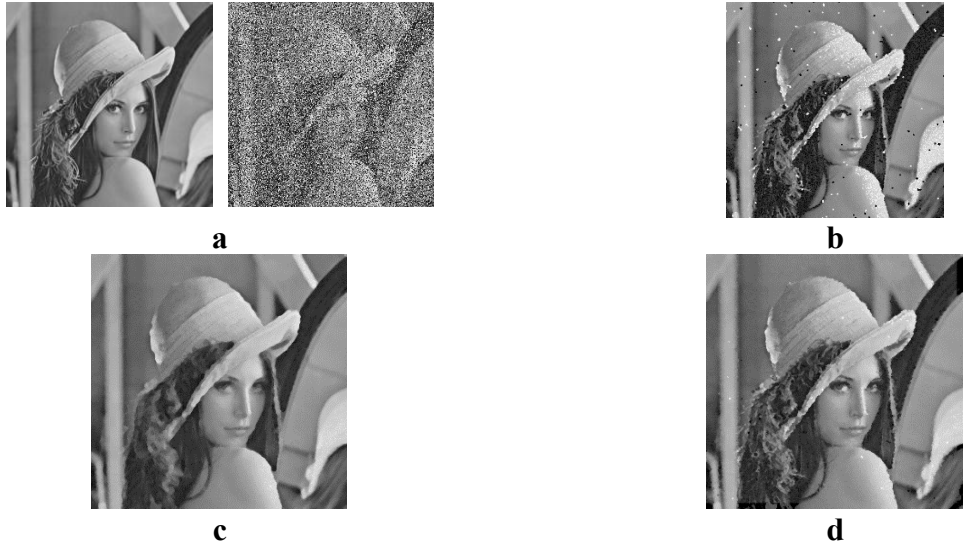

**Fig. 12.** (a) Clean (left) and heavily SP corrupted (right) *Lena* image with $\mu_{sp}$ = 61 %. (b), (c), and (d) Restored *Lena* images obtained using MF with (3×3) window, with (5×5) window, and the MFs-AE scheme, respectively. (b) $SSIM_{Img}$ = 0.80; (c) $SSIM_{Img}$ = 0.82; (d) $SSIM_{Img}$ = 0.83. Calculation parameters are in Table 3.

As seen in Fig. 12b (top), MF with (3×3) window leaves small "islands" of artifacts (black and white speckles), similar to Fig. 8a (bottom). The MFs-AE scheme (Fig. 12d) yields the best restoration, which becomes clear only on the entropy maps (Fig. 13). The relative difference $\Delta SSIM_{Map}$ (5) between the MF with (5×5) window (Fig. 13c) and the MFs-AE scheme (Fig. 13d) is ~3.5% for the standard maps (left) and is ~8% for the dilation-enhanced copies (right).

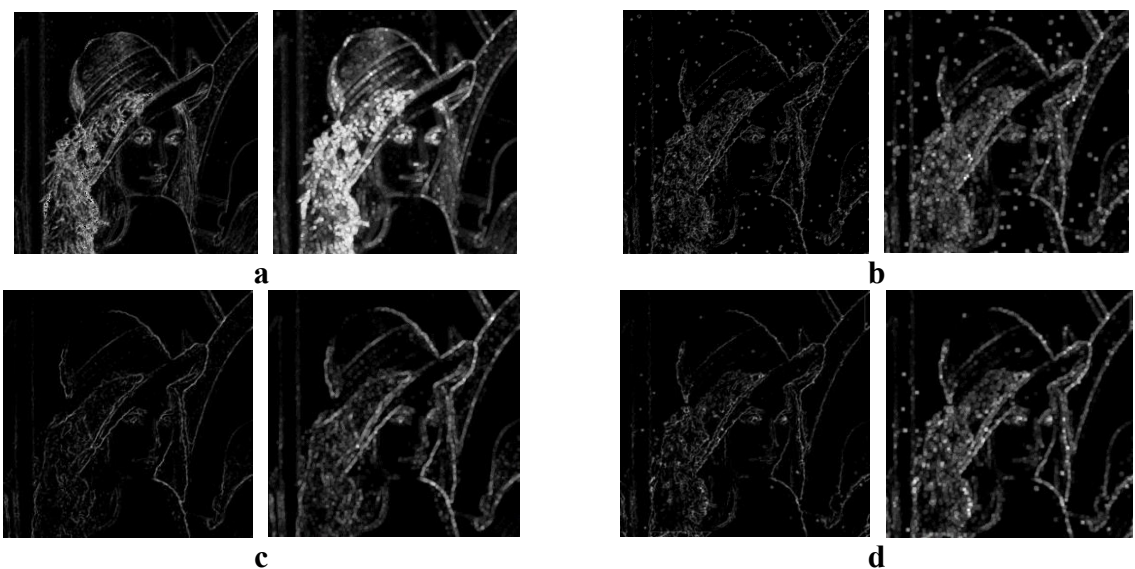

**Fig. 13.** Entropy maps of the clean (a) and restored, heavily corrupted ($\mu_{sp}$ = 61 %). *Lena* images obtained with MF using (3×3) window (b), (5×5) window (c), and the MFs-AE scheme (d). Left: standard entropy maps of the images. Right: copies after morphological dilation (kernel (5×5), one iteration). (b) $SSIM_{Map}$ = 0.42 (left), $SSIM_{Map}$ = 0.34 (right); (c) $SSIM_{Map}$ = 0.43 (left), $SSIM_{Map}$ = 0.36 (right); (d) $SSIM_{Map}$ = 0.45 (left), $SSIM_{Map}$ = 0.38 (right).

**Table 4** A comparison of $SSIM_{Img}$ across different methods

| Methods | DAF [24] | DBA [25] | NAFSM [26] | INLMF [27] | DAMF [28] | FSAP [29] | DCNN [30] | This work | | |
|---|---|---|---|---|---|---|---|---|---|---|
| | | | | | | | | MF (5×5) | 2MF | MFs-AE |
| SSIM | 0.89 | 0.82 | 0.87 | 0.77 | 0.87 | 0.76 | 0.93 | 0.84 | 0.85 | 0.87 |

\* Methods: detail-aware filter (DAF), decision-based algorithm (DBA), noise adaptive fuzzy switching median filter (NAFSM), iterative nonlocal mean filter (INLMF), different applied median filter (DAMF), fuzzy approach for salt-and-pepper noise (FSAP), deep convolutional neural network (DCNN). MF-based scheme parameters for "This work" are in Table 3.

Finally, Table 4 reports denoising results for the grayscale *Lena* image (512×512 px) with fixed SP noise ($\mu_{sp}$ = 61 %) for methods [24–30], compared with the algorithms proposed in this study.

# 4 Discussion

The simulations in Section 3.1 corroborate a known result [8,9]: as the number of recursions increases, the MF converges to a stationary root image that no longer changes and achieves maximal $SSIM_{Img}$ and $SSIM_{Map}$ In contrast, the recursing and thresholding for the AE have only a minor effect and mainly at low noise levels. Overall, AE can compete with MF only under mild SP noise; it performs substantially worse for heavily corrupted images under recursive threshold denoising. As a simple three-layer network, the AE cannot reliably learn image features when the background is excessively noisy; it requires preprocessing, e.g., noise reduction by the MF, after which the AE can further refine the restoration.

A common pattern in the proposed 2MF and MFs-AE schemes is the use of MFs at the first stage and a final thresholding step. Step 1 produces two types of outputs: (i) one (in 2MF) or a set (in MFs-AE) of images that still contain "islands" of impulse artifacts, and (ii) a single, more blurred image largely free of impulse corruption. At the final step, the thresholding effectively replaces corrupted pixels in the first-type images with the corresponding pixels from the second, blurrier image. The final output is therefore less blurred than the second type yet free of SP noise like the second type.

The autoencoder's role in the MFs-AE scheme is less about additional denoising and more about fusing common features across the first-type inputs. Because these inputs are generated by MFs with different thresholds, they exhibit different contrast distributions and artifact "islands." The AE extracts an aggregate image with averaged contrast (Step 2), and the thresholding (Step 3) removes the residual islands.

The comparison in Table 4 shows that the recursive median filtering with thresholding studied here, especially the multi-level MFs-AE scheme, can compete with several modern denoising approaches [24–30]. In terms of $SSIM_{Img}$ MFs-AE trails mainly the more complex methods such as DAF [24] (bilateral-filter–based restoration) and DCNN [30] (deep learning). We also note that fixed SP noise is suppressed slightly more effectively than the interval SP model in our schemes, as seen by comparing the results in Table 4 with the $SSIM_{Img}$ values in Fig. 12c and 12d.

There are many entropy measures, including 2D variants, for building entropy maps, and they differ in segmentation behavior and textural detail. Depending on the entropy type and parameters, "entropy images" can be highly or weakly detailed, reflecting the metric's sensitivity to local intensity changes. We focused on SampEn, which, for our parameters, yields maps with the clearest delineation of key structures: the mushroom cap (Fig. 3b) and the face in *Lena* (Fig. 13a). Simulations indicated that alternative entropy measures often provided weaker or ambiguous blur assessment, and in some cases were numerically unstable. Moreover, as resolution increases, entropy-map contrast tends to decrease; thus we applied morphological dilation to enhance contrast and computed SSIM on the modified maps as well. Comparing Fig. 13c and 13d shows that the relative difference $\Delta SSIM_{Map}$ (5) for the dilated maps is more than twice that for the standard maps. Hence, post-processing of "entropy images" (e.g., dilation) can increase SSIM sensitivity and yield a more reliable contrast assessment.

## 5 Conclusion

Using SSIM metrics, we showed that recursive threshold median filtering reliably restores images under strong SP noise, even when the target structure is barely visible. The MF consistently converges to a stationary "root" image and preserves edges via thresholded substitution; at extreme noise levels, gains appear after the very first recursions. The two proposed scheme (2MF and MFs-AE) extend MF's capabilities: 2MF is better suited to low-resolution images and tasks emphasizing sharp local details (edges), whereas MFs-AE provides a modest yet reproducible improvement on higher-resolution images by aggregating features across multiple MFs followed by thresholding. MF remains unmatched for deployment on resource-constrained platforms (e.g., mobile IoT sensors) due to its simplicity and efficiency; without prior denoising, AE underperforms and requires preprocessing.

Our SSIM measure for entropy maps complements the standard $SSIM_{Img}$ and better reflects the presence/absence of sharp details. Its sensitivity increases with simple post-processing of entropy maps (dilation), making $SSIM_{Map}$ a practical tool for parameter tuning (window sizes, thresholds, recursion count) and objective blur assessment.

Further research can be conducted in both directions: (i) adaptive thresholds and multiscale windows, MF↔AE hybrids with learned fusion, extensions to mixed-noise and color images; (ii) broader entropy-based analysis (alternative 2D entropies, scale/contrast-robust metrics, other morphological operators), expanding the scope of application (recognition, classification, compression, coding, etc.), and hardware implementations for edge devices.

**Funding** This research was supported by the Russian Science Foundation (grant no. 22-11-00055-P, https://rscf.ru/en/project/22-11-00055/, accessed on 10 June 2025).

**Authors Contributions** Petr Boriskov, conceptualization and methodology; Petr Boriskov, and Kirill Rudkovskii, software and investigation; Petr Boriskov, and Andrey Velichko, revised and finalized the paper; Andrey Velichko, drafted the manuscript and project administration.

**Conflicts of interest/Competing interests**
Petr Boriskov, Kirill Rudkovskii, and Andrey Velichko declare that they have no conflict of interest.